\documentclass[a4paper, 11pt]{article} 
\usepackage{amsmath}
\usepackage{authblk}
\usepackage[margin=1 in]{geometry}

\begin{document}

\title{\huge Eigenlogic and Probabilistic Inference;\\when Bayes meets Born{\footnote {\rm  \small $\,$ This contribution 
    is  published in {\it Philosophical Transactions of the Royal Society A},
    volume 383, issue 2310, article 20240392 [15 pages], 11 December 2025.}}}

\author[$1,2$]{\Large François Dubois}
\author[$3$]{Zeno Toffano}

\affil[$1$]{\small Conservatoire National des Arts et Métiers, LMSSC laboratory,  Paris, France.}
\affil[$2$]{Association Française de Science des Systèmes, Paris, France.}
\affil[$3$]{Laboratoire Signaux et Systèmes (L2S), UMR 8506, CentraleSupélec, Université Paris-Saclay, CNRS, 91190 Gif-sur-Yvette, France.}

\date{February 7, 2025}

\maketitle

\begin{abstract}
This paper shows how inference is treated within the context of Eigenlogic projection operators in linear algebra. In Eigenlogic operators represent logical connectives, their eigenvalues the truth-values and the associated eigenvectors the logical models.
By extension, a probabilistic interpretation is proposed using vectors outside the eigensystem of the Eigenlogic operators. The probability is calculated by the quantum mean value (Born rule) of the logical projection operators.
We look here for possible connections between the Born rule in quantum mechanics and Bayes' theorem from probability theory and show that Eigenlogic offers an innovative approach to address the probabilistic version of logical inference (material implication) in a quantum context.
\end{abstract}

\section{Introduction}

Inference is important in statistics, machine learning, and artificial intelligence, where Bayes' theorem is commonly used to update probabilities based on new evidence. In quantum mechanics, the quantum measurement outcome probability is given by Born's rule. Both frameworks aim to assign likelihoods to possible outcomes.

In recent years, the \textit{Eigenlogic} method \cite{EL-LNCS 2016, EL-LU 2020} has been proposed to formalize logic within a quantum context. Eigenlogic is a quantum-like formalism that uses operators in a Hilbert space to represent logical propositions with the eigenvalues representing the truth values and the eigenvectors the logical models.

This formulation also allows for a probabilistic interpretation compatible with quantum mechanics. The use of operators for logic has interesting implications for quantum computing and information; several applications of Eigenlogic have been proposed \cite{EL-Proceed 2018,EL-Entropy 2020}.

Eigenlogic provides an interpretation of logical inference that is both probabilistic and quantum. In this paper, we will confront Bayes' theorem and Born's rule in this framework, proposing a probabilistic inference model.
 
\section{Inference and Probability} \label{Probabilistic Inference}

Bayes' theorem \cite{Bayes 1763} is fundamental to classical probability theory and provides a way to update probabilities based on evidence. In the classical world, the conditional probability of event \(B\) given event \(A\) is given by:

\begin{align}
\label{Bayes rule}
\begin{split}
P(B \mid A) = \frac{P(A \cap B)}{P(A)}=\frac{P(A \mid B)P(B)}{P(A)},
\end{split}
\end{align}

where \( P(B|A) \) is the posterior probability, \( P(A|B) \) is the likelihood, \( P(B) \) is the prior probability, and \( P(A) \) is the marginal likelihood.

Bayesian probabilities are generally considered the correct way to deal with inference, but difficulties appear when dealing with logical inferences such as \textit{modus ponens} and \textit{modus tollens}. A probabilistic version of logical inference, usually called \textit{material implication}, could become necessary when information about prior or conditional probabilities is not available \cite{Nguyen 2002,Jahn Scheutz 2023}.

Boole in \cite{Boole 1854}  already combined probability and logical inference, quoting Hailperin in \cite{Hailperin 1984} : "one of the examples Boole works with is that of finding the probability of the conclusion of a hypothetical syllogism".  We will discuss the methods developed by Boole in the next section.

Probabilities can be associated with the usual logical propositional connectives by the standard principles of probability theory, where set-theoretic operators enable probabilistic interpretations of propositional sentences, with the correspondence of \textit{intersection} \(\cap\) with \textit{conjunction} \(\land\) (AND) and \textit{union} \(\cup\) with \textit{disjunction} \(\lor\) (OR).
For example, we have the well-known property :

\begin{align}\label{prob disjunction}
\begin{split}
P(A\cup B)~~=~~P(A)+P(B)-P(A\cap B),
\end{split}
\end{align}
\\
which, as we will see in the following section \ref{Logical Inference and Eigenlogic}, has a logical version. 

When considering material implication \((A \rightarrow B)\) the sentence "if \(A\) then \(B\) is true" means practically that whenever \(A\) is true, \(B\) is true as well; classically this is named \textit{modus ponens}. The first statement (“if…”) is called the antecedent, and the second (“then...”) is called the consequent.
The material implication is false only when the antecedent is true and the consequent is false, and in all other cases it is always true (see the corresponding truth table in table \ref{Truth table}). The consequence of this definition is that if there is no relation between the antecedent and the consequent, then the truth-functional definition of material implication leads to a paradox. For example, 'if the earth is flat, then the sun is a star' and 'if the earth is flat then the sun is not a star' are true sentences for material implication because the antecedent is false.
In practical situations the cases where the antecedent is false in material implication are not considered  and are still subject to much debate .

In real situations, we are generally not sure of the truth of \(A\), alternatively we can have uncertainty on the veracity of the implication \((A \rightarrow B)\), in these cases we have to look for estimates of probabilities of \(P(A)\) and \(P(A \rightarrow B)\).
The probabilistic version of the inference rule ("if \(A\) then \(B\) ") have either a Bayesian interpretation via conditional probabilities \(P(B|A)\), as given by equation (\ref{Bayes rule}), or can be expressed with probabilistic material implication forms \(P(A \rightarrow B)\) as we will show hereafter.
Probabilities associated to logical inference were investigated in \cite{Wagner 2004} where a probabilistic version of the \textit{modus ponens} inference rule was proposed.

Using the logical transformation (see the corresponding truth tables in table \ref{Truth table} in the next section \ref{Logical Inference and Eigenlogic}):

\begin{align}\label{mat implic tranform}
\begin{split}
A \rightarrow B=\neg A \lor B=\neg A\lor (A\land B),
\end{split}
\end{align}
\\
one can obtain an expression of the probability of logical material implication:

\begin{align}\label{proba mat implic}
\begin{split}
P(A \rightarrow B)=P(\neg A)+P(A \land B)=1-P(A)+P(A \cap B).
\end{split}
\end{align}
\\
Equation (\ref{proba mat implic}) can also be expressed as:

\begin{align}\label{proba mat implic 2}
\begin{split}
P(A \cap B)=P(A)-1+P(A \rightarrow B).
\end{split}
\end{align}
\\

There is a symmetric form with converse implication  \((B \rightarrow A)\):

\begin{align}\label{proba mat implic 3}
\begin{split}
P(A \cap B)=P(B)-1+P(B \rightarrow A).
\end{split}
\end{align}
\\
From the preceding two equations (\ref{proba mat implic 2}) and (\ref{proba mat implic 3}) one obtains an expression that relates the probability of implication and the probability of the converse implication.

\begin{align}\label{proba mat implic 4}
\begin{split}
P(A)+P(A \rightarrow B)=P(B)+P(B \rightarrow A)=P(A \cap B)+1 .
\end{split}
\end{align}

The preceding expression (\ref{proba mat implic 4}) is interesting because it provides a symmetric relationship between probabilities of events and
their implications and could represent a linear alternative to Bayes' rule (\ref{Bayes rule}) for the cases where we can interpret the posterior conditional probability \( P(B|A) \) as \( P(A \rightarrow B) \) and the likelihood conditional probability \( P(A|B) \) as \( P(B \rightarrow A)\).

Several authors \cite{Nguyen 2002,Wagner 2004,Jahn Scheutz 2023} have tackled the probabilistic version of material implication. In \cite{Nguyen 2002} and \cite{Jahn Scheutz 2023} the authors derived expressions that show the different situations where formulas do coincide and do not coincide with the Bayes' rule. Essentially, an interpolation formula is derived \cite{Jahn Scheutz 2023} that bridges the two cases (material implication and Bayes rule) :

\begin{align}\label{proba Bayes-MatImplic}
\begin{split}
P(A \overset{\alpha}{{\rightarrow}} B)=\frac{P(A \land B)+\alpha P(\neg A)}{P(A)~+~~\alpha P(\neg A)}=\frac{P(A \cap B)+\alpha (1-P(A))}{P(A)~+~~\alpha (1-P(A))},
\end{split}
\end{align}
\\
depending on the coefficient  \(\alpha\)
we have for \(\alpha=0\) the Bayesian case according to equation (\ref{Bayes rule}) and for \(\alpha=1\) the material implication case according to equation (\ref{proba mat implic}).
\\

Another objective of a probabilistic version of logical propositions is to find probability bounds \cite{Jokismovic 2021}.  Bounds related to conjunction and disjunction can be derived from \textit{Boole's inequality} and \textit{Bonferroni's inequality}, as we will show hereafter.

Starting from the relation (\ref{prob disjunction}), due to the fact that probabilities must be positive, one has the inequality: 

\begin{align}\label{prob disjunction inequality}
\begin{split}
P(A\cup B)~~~ \leq ~~P(A)+P(B) .
\end{split}
\end{align}

Often in a probabilistic setting, following the axioms of Kolmogorov's probability theory, one only considers independent events, so $P(A\cap B)=0$ and the preceding inequality (\ref{prob disjunction inequality}) becomes an equality.

Again using equation (\ref{prob disjunction}) by increasing the number of disjunctive propositions, one has the well-known \textit{inclusion-exclusion principle} outlined by Henri Poincaré in \cite{Poincare-1912}.

\begin{align}\label{Incl Excl}
\begin{split}
P(\cup_{i=1}^{n}A_i)=\sum_{i=1}^{n} P(A_i)-\sum_{i,j=1, i<j}^{n} P(A_i\cap A_j)+\sum_{i,j,k=1, i<j<k}^{n} P(A_i\cap A_j\cap A_k))-...
\end{split}
\end{align}

From the above expression, one obtains what is named the \textit{Boole inequality}:

\begin{align}\label{Boole Ineq}
\begin{split}
P(\cup_{i=1}^{n}A_i)~~\leq~~\sum_{i=1}^{n} P(A_i).
\end{split}
\end{align}
\\

By a similar reasoning one also obtains the \textit{Bonferroni’s inequality}:

\begin{align}\label{Bonferroni Ineq}
\begin{split}
P(\cup_{i=1}^{n}A_i)~~\geq~~\sum_{i=1}^{n} P(A_i)-\sum_{i,j=1, i<j}^{n} P(A_i\cap A_j).
\end{split}
\end{align}
\\

Inference probability bounds are more involved, from equations (\ref{mat implic tranform}) and (\ref{Boole Ineq}) we obtain the following inequalities:

\begin{align}\label{Implic Ineq 1}
\begin{split}
P(A \rightarrow B)=P(\neg A \lor B)~~\leq~~ P(\neg A) + P(B)=1-P(A)+P(B)
\end{split}
\end{align}
\\
and
\begin{align}\label{Implic Ineq 2}
\begin{split}
P(B)=P(A\land B)+P(\neg A \land B)~~\leq~~ P(A\land B) + P(\neg A)=P(A \rightarrow B).
\end{split}
\end{align}
\\

The question arises now: is there an equivalence between a probabilistic version of material implication and the Bayes' rule when applied in a probabilistic context on one side and a logic or quantum context on the other? 

In the following section \ref{Logical Inference and Eigenlogic} we will derive Bayes-like formulas for logic, using Boole's method in (\ref{logic Bayes}) and using Eigenlogic in (\ref{EL Bayes}) and also a quantum-like version having a link to Born's rule in section \ref{Quantum Born and Bayes' rule} (see equation (\ref{Quantum Bayes'})) and in section \ref{Quantum-like Bayes' conditions} (see equations (\ref{Bayes=Prob-Implication}) and (\ref{Bayes=Prob-RevImplication})).

In the framework of Eigenlogic we will discuss the quantum situation where the projection operator corresponding to logical implication \( A \rightarrow B \) can be understood as a conditional projection representing the conditional probability.

\section{Logical Inference and Eigenlogic} \label{Logical Inference and Eigenlogic}

Logical inference, also called \textit{material implication}, is defined by the logical connective \( A \rightarrow B \), involving two elementary propositions \( A \) and \( B \), where the implication is false only if \( A \) is true and \( B \) is false.

The operation is best illustrated by its truth table for the four possible combinations (the \textit{logical models}) of the two elementary propositions \( A \) and \( B \).
Using truth values 1 for True and 0 for False, the truth tables for the connectives of logical conjunction \( A \land B\),  disjunction \( A \lor B\), implication \( A \rightarrow B \), reverse implication \( B \rightarrow A \) and the equivalent disjunctive forms \( \neg A \lor B \) and \( A \lor \neg B \) are shown in table \ref{Truth table}:

\begin{table}[!h]
\caption{Truth tables for: \( A \), \( B \), their negations \( \neg A\) , \(\neg B\), conjunction  \(A \land B\) (AND), disjunction  \(A \lor B\) (OR), implication \( A \rightarrow B \), reverse implication \( B \rightarrow A \) and their equivalent implication disjunction forms \( \neg A \lor B \) and \( A \lor \neg B \) .}
\label{Truth table}
\begin{tabular}{|l|l|l|l|l|l|l|l|l|l|}
 \hline  
\( A \) &\( B \)&\( A \land B\) &\( A \lor B\) &\( A \rightarrow B \)&\( B \rightarrow A \)&\( \neg A \)&\( \neg B \)&\( \neg A \lor B\)&\( A \lor \neg B\)\\ \hline  

0 & 0 & 0 & 0 & 1 & 1 & 1 & 1 & 1 & 1 \\ \hline  
0 & 1 & 0 & 1 & 1 & 0 & 1 & 0 & 1 & 0 \\ \hline  
1 & 0 & 0 & 1 & 0 & 1 & 0 & 1 & 0 & 1 \\ \hline  
1 & 1 & 1 & 1 & 1& 1 & 0 & 0 & 1 & 1 \\ \hline 
\end{tabular}
\vspace*{-4pt}
\end{table}

These tables clarify the logical structure and show some important properties for the implication connective:
\begin{itemize}

    \item The implication connective \( A \rightarrow B \)  has an equivalent disjunctive form \( \neg A \lor B\).
    \\
     \item The reverse implication connective  \( B \rightarrow A \)  has an equivalent disjunctive form \( A \lor \neg B\).
     \\
    \item Considering the truth values \( tv \) (0 or 1) of disjunction \( A \land B\) , implication \( A \rightarrow B \)  and reverse implication \( B \rightarrow A \) in the table \ref{Truth table} one gets a relation having the same structure as Bayes' rule for probabilities given in equation (\ref{Bayes rule}) :
    \begin{align}\label{logic Bayes}
\begin{split}
tv(A)\cdot tv(A \rightarrow B)~=~ tv(B)\cdot tv(B \rightarrow A)~=~tv(A\land B).
\end{split}
\end{align}  
\item Another expression involving the truth values  \( tv \) (0 or 1)  has the same form as equation (\ref{proba mat implic 4}) involving  probabilities:
\begin{align}\label{logic Bayes sum}
\begin{split}
tv(A)\ + tv(A \rightarrow B)~=~ tv(B)\ + tv(B \rightarrow A)~=1+~tv(A\land B).
\end{split}
\end{align} 
\end{itemize}

This original method was proposed by George Boole for logic in 1847 \cite{Boole 1847} and was successively extended to probabilities in 1854 \cite{Boole 1854} as is thoroughly discussed in \cite{EL-LU 2020,EL-Entropy 2020,Hailperin 1984}.

Technically, the method uses idempotent symbols taking the values 0 and 1 for logical propositions and then arithmetic operations to manipulate and transform them. The idempotence law represented by the equation \( x^2= x \) was defined by Boole as the "fundamental law of thought" and allowed him to give a mathematical binary formulation of logic using the two numbers 0 and 1 which are the only possible solutions of the idempotence equation.

The power of the method lies in the fact that knowing the truth values of a logical connective one can obtain the respective arithmetic logical function by interpolation.
Considering a logical function \( f_L\), corresponding to the logical connective \( L \) of the two arguments \( A \) and \(B \),  represented by the idempotent symbols \( a= tv(A) \) and \( b= tv(B) \), with \( a^2= a \) and \( b^2= b \), one simply writes, as was originally proposed in \cite{Boole 1847}, the logical function using the four interpolated truth values $\{f_L(0,0), f_L(0,1), f_L(1,0), f_L(1,1)\}$ :

\begin{align}\label{F interpol}
\begin{split}
f_L(a,b)~=~(1-a)(1-b)~f_L(0,0)+(1-a)b~f_L(0,1)+a(1-b)~f_L(1,0)+ab~f_L(1,1),
\end{split}
\end{align}
\\
where the complement \( 1-a \) represents logical negation \( \neg A \).

Using the values of the truth table \ref{Truth table} and the interpolation equation (\ref{F interpol}), the function for conjunction ($L=A\land B$) is simply given by the product because here only one truth value,  $f_L(1,1)$,  is equal to $1$:

\begin{align}\label{conjunction}
\begin{split}
f_{A\land B}~=~ab  .
\end{split}
\end{align}  
\\

The disjunction function ($L=A\lor B$) is more involved because we have three truth values that are equal to $1$ : $f_L(0,0)=0$ and $f_L(0,1)=f_L(1,0)=f_L(1,1)=1$

\begin{align}\label{disjunction}
\begin{split}
f_{A\lor B}~=(1-a)b+a(1-b)+a b=~a~+~b~-~a b.
\end{split}
\end{align}
\\
Here we find the same structure as in the probabilistic formula (\ref{prob disjunction inequality}) for $P(A\cup B)$.

Using the same method, the logical function for material implication ($L=A\rightarrow B$), using $f_L(1,0)=0$ and $f_L(0,0)=f_L(0,1)=f_L(1,1)=1$, is :

\begin{align}\label{implication}
\begin{split}
f_{A \rightarrow B}~=~(1-a)(1-b)+(1-a)b+a b=~1~-~a~+~a b.
\end{split}
\end{align}  
\\

It is now straightforward (using the idempotence of the symbols) to justify the Bayes rule-like result above (\ref{logic Bayes})  derived from the truth tables:

\begin{align}\label{logic Bayes idemp}
\begin{split}
f_A  f_{A \rightarrow B}=a(1-a+a b)=a b=f_{A\land B}=b(1-b+a b)=f_B  f_{B \rightarrow A},
\end{split}
\end{align}
\\
and also to justify the result (\ref{logic Bayes sum}):

\begin{align}\label{logic Bayes idemp sum}
\begin{split}
f_A \ + f_{A \rightarrow B}=a+(1-a+a b)=1+a b=1+f_{A\land B}=b+(1-b+a b)=f_B \ + f_{B \rightarrow A}.
\end{split}
\end{align}
\\

Birkhoff and Von Neumann formalized in 1936 \textit{quantum logic} as a lattice of projectors \cite{Birkhoff Neumann 1936} and successively many mathematical constructions were developed to extend and improve this formulation \cite{Dalla Chiara et al. 2013} . But since the beginning it appeared that none of these constructions brought a solid basis for a logical system that can be used for reasoning; essentially because these logical constructions have problems with the logical mechanisms of implication.

We introduced Eigenlogic in 2016\cite{EL-LNCS 2016}. Eigenlogic extends Boole's logical method by representing propositions as projection operators, which are also idempotent.

In Eigenlogic logical operations are defined in terms of linear algebra. The logical connectives (AND, OR...) are represented by operators, and more specifically here projection operators.

All logical propositions generated by the different logical connectives are represented in Eigenlogic by commuting projection operators, and belong to the same Eigenlogic family  with the same eigenvectors (the same \textit{context}). It is well known that if we limit ourselves to commutative projections, we get the entire usual classical logic, this is also justified by Boole's method.

The truth values of a logical proposition, 1 (true) and 0 (false), are the possible eigenvalues of the respective Eigenlogic operator.

The eigenvectors associated to every eigenvalue (truth value) of the Eigenlogic operator correspond to what in logic is named a \textit{logical model}, one of the possible combinations of the logical input values (called the \textit{elementary propositions}). In truth tables, the logical models correspond to the different lines of the table.

In the case of 2 logical inputs (\textit{arity}-2) one has four possible logical models which represent all the possible combinations of the inputs and correspond, in the language of quantum information, to the 4 eigenvectors: $|00\rangle$, $|01\rangle$, $|10\rangle$ and $|11\rangle$ which are the 2-qubit computational basis states.

All the logical models of a logical proposition form what is called a \textit{logical interpretation}, this represents the \textit{context} in a quantum perspective.

The interesting fact is that Eigenlogic permits to go beyond the classical logic cases described above when one considers different contexts \textit{i.e.} vectors outside the eigensystem. By applying the Eigenlogic operators outside the given context one introduces indeterminacy.  By taking the quantum mean values of the Eigenlogic operators, one gets the probability associated to the considered logical proposition in a new context. This consists essentially in the application of the quantum \textit{Born rule}  \cite{Born 1926} as we will show hereafter.

All logical operators can be expressed, using Boole's method shown above, by replacing the symbols \( a \) and \( b \) by the respective commuting projectors \(\boldsymbol{A}\) and \(\boldsymbol{B}\) which verify the idempotence and commutativity conditions:

\begin{align}\label{idemp commut ooerators}
\begin{split}
\boldsymbol{A}^2 = \boldsymbol{A},~~~~ \boldsymbol{B}^2 = \boldsymbol{B},~~~~ \boldsymbol{A}\boldsymbol{B} = \boldsymbol{B} \boldsymbol{A}.
\end{split}
\end{align}
\\

The operator \(\boldsymbol{F}_{A\land B}\) corresponding to the logical conjunction (AND) is the \textit{intersection} $\cap$ of the projectors \(\boldsymbol{A}\) and \(\boldsymbol{B}\), simply given by the product of the projectors as in (\ref{conjunction}):

\begin{align}\label{operator AND}
\begin{split}
\boldsymbol{F}_{A\land B} = \boldsymbol{A} \boldsymbol{B}.
\end{split}
\end{align}
\\

The projection operator \( \boldsymbol{F}_{A\lor B}\) corresponding to logical disjunction (OR) is the \textit{union} $\cup$ of \(\boldsymbol{A}\) and \(\boldsymbol{B}\) and given by an expression analogue to (\ref{disjunction}):

\begin{align}\label{operator AND}
\begin{split}
\boldsymbol{F}_{A\lor B} =\boldsymbol{A}+\boldsymbol{B}-\boldsymbol{A} \boldsymbol{B}.
\end{split}
\end{align}

In the same way, the material implication operator analogue to (\ref{implication}) is defined as :

\begin{align}\label{operator inference}
\begin{split}
\boldsymbol{F}_{A \rightarrow B}  = \boldsymbol{I} - \boldsymbol{A} + \boldsymbol{A} \boldsymbol{B},
\end{split}
\end{align}
\\
where \( \boldsymbol{I} \) is the identity operator.
Because operators commute we have :

\begin{align}\label{commut Bayes}
\begin{split}
\boldsymbol{A}\boldsymbol{F}_{A \rightarrow B}=\boldsymbol{F}_{A \rightarrow B}~\boldsymbol{A}.
\end{split}
\end{align}

We then use equation (\ref{logic Bayes idemp}) to express the Bayes-like rule in Eigenlogic:

\begin{align}\label{EL Bayes}
\begin{split}
\boldsymbol{A}\boldsymbol{F}_{A \rightarrow B}  = \boldsymbol{A}  (\boldsymbol{I} - \boldsymbol{A} + \boldsymbol{A} \boldsymbol{B}  ) = \boldsymbol{A} - \boldsymbol{A}^2 + \boldsymbol{A}^2 \boldsymbol{B}=\boldsymbol{A} \boldsymbol{B} =  \boldsymbol{F}_{A\land B} =\boldsymbol{B}\boldsymbol{F}_{B \rightarrow A}=\boldsymbol{B}\boldsymbol{A}
\end{split}
\end{align}
\\
this because \( \boldsymbol{A}^2 = \boldsymbol{A}\),  \( \boldsymbol{B}^2 = \boldsymbol{B}\) and \(\boldsymbol{A}\boldsymbol{B}=\boldsymbol{B}\boldsymbol{A}\).

Eigenlogic uses the structure of Hilbert space and projection operators to represent and manipulate logical propositions. When one considers general quantum states not belonging to the Eigenlogic eigensystem one obtains a probabilistic representation.

As an example of a general quantum state $|\psi\rangle$, let us consider a separable two-qubit state obtained by taking the tensor product of the two one-qubit states $|\varphi_{p}\rangle$ and $|\varphi_{q}\rangle$:

\begin{align}\label{2-qubit p q}
\begin{split}
\begin{array} {l}
|\psi\rangle=|\varphi_{p}\rangle\otimes|\varphi_{q}\rangle,
\\[3pt]
|\varphi_{p} \!\rangle = \cos\frac{\theta_{p}}{2}|0\rangle+e^{i\varphi_{p}}\sin\frac{\theta_{p}}{2}|1\rangle,
\\[3pt]
|\varphi_{q} \!\rangle = \cos\frac{\theta_{q}}{2}|0\rangle+e^{i\varphi_{q}}\sin\frac{\theta_{q}}{2}|1\rangle,
\end{array}
\end{split}
\end{align}
\\
where $\sin^{2}\frac{\theta_{p}}{2}=p$
and $\sin^{2}\frac{\theta_{q}}{2}=q$ are the probabilities of being
in the state $|1\rangle$ (which represents in Eigenlogic the logical "true" state) for two spins $\frac{1}{2}$ oriented
along two different axes $\theta_{p}$ and $\theta_{q}$ .

The probabilities associated with the logical projectors $\boldsymbol{A}$ and $\boldsymbol{B}$ using Equation (\ref{2-qubit p q}) and the \textit{Born rule} \cite{Born 1926} are given by:
\\
\begin{align}\label{membership A B}
\begin{split}
\begin{array} {l}
\langle\psi|\boldsymbol{A}|\psi\rangle =
p(1-q)+pq=p,
\\[3pt]
\langle\psi|\boldsymbol{B}|\psi\rangle=(1-p)q+p q=q.
\end{array}
\end{split}
\end{align}
\\
This means that the quantum mean values of the Eigenlogic operators correspond to the respective probabilities.

Now one can "measure" also the conjunction $A\land B$ , the disjunction $A\lor B$  and material implication  $A\rightarrow B$ using the respective logical projectors:

\begin{align}\label{fuzzy AND OR}
\begin{split}
\begin{array} {l}
\langle\psi|\boldsymbol{F}_{A\land B}|\psi\rangle~=~\langle\psi|\boldsymbol{A}\boldsymbol{B}|\psi\rangle~=~  p  q ,
\\[3pt]
\langle\psi|\boldsymbol{F}_{A\lor B}|\psi\rangle~=~ p+q-p q,
\\[3pt]
\langle\psi|\boldsymbol{F}_{A\rightarrow B}|\psi\rangle~=~ 1-p+p q.
\end{array}
\end{split}
\end{align}

So for a general quantum state-vector, 
$|\psi\rangle$ (separable or entangled), the mean value of whatever logical projector of the type $\boldsymbol{F}$ is given, by the \textit{Born rule} \cite{Born 1926} and will always verify the inequality:

\begin{align}\label{Born Gleason}
\begin{split}
\langle\psi|\boldsymbol{F}|\psi\rangle  \,\,=\,\,
\mathrm{Tr} \, (\rho_{\Psi}\cdot\boldsymbol{F}) \,\leq\, 1 \,, \qquad {\rm with} 
\quad \rho_{\Psi} \,\, \equiv \,\, |\psi\rangle \, \langle\psi|,
\end{split}
\end{align}

and can thus be understood as a probability measure.

\section{Quantum Born and Bayes' Rule} \label{Quantum Born and Bayes' rule}

In the framework of quantum mechanics, a general quantum state vector \( |\psi\rangle \)  represents the context of the system. Probabilities are calculated by taking the mean value of the projection operators on the considered state vectors, yielding quantum probabilities  \( P_\psi \) according to \textit{Born's rule} \cite {Born 1926}, as shown in the preceding section \ref{Logical Inference and Eigenlogic}, this gives using Eigenlogic: 

\begin{align}\label{commut Bayes}
\begin{split}
 \begin{array} {l}
  P_\psi(A) = \langle\psi | \boldsymbol{A} | \psi\rangle,
\\[3pt]
  P_\psi(B) = \langle\psi | \boldsymbol{B} | \psi\rangle,
\\[3pt]
  P_\psi(A\cap B) = \langle\psi | \boldsymbol{F}_{A\land B} | \psi\rangle = \langle\psi | \boldsymbol{A} \boldsymbol{B} | \psi\rangle,
\\[3pt]
  P_\psi(A\cup B) = \langle\psi | \boldsymbol{F}_{A\lor B} | \psi\rangle = \langle\psi | \boldsymbol{A} | \psi\rangle + \langle\psi | \boldsymbol{B} | \psi\rangle - \langle\psi | \boldsymbol{A} \boldsymbol{B} | \psi\rangle,
\\[3pt]
P_\psi(A\rightarrow B) = \langle\psi | \boldsymbol{F}_{A \rightarrow B} | \psi\rangle = 1-\langle\psi | \boldsymbol{A} | \psi\rangle + \langle\psi | \boldsymbol{A} \boldsymbol{B} | \psi\rangle.
\end{array}  
\end{split}
\end{align}
\\

The preceding expressions show that one can get probabilities linked to logical propositions and in particular logical inference by applying the Born rule depending on the context of the system under consideration. 

A natural question related to inference arises: is it possible to identify the quantum probability for material implication $ \, P_\psi(A\rightarrow B) \, $
with the quantum conditional probability $ \, P_\psi(B | A) $ of $ \, B \, $  knowing $ \, A $? Can we write the following equality?

\begin{align}\label{Quantum Bayes'}
\begin{split}
P_\psi(B | A) \equiv   {{ P_\psi(A\cap B)}\over{P_\psi(A)}}   =  P_\psi(A\rightarrow B) \,  ?
\end{split}
\end{align}
\\
Let us define a general quantum state orthogonal decomposition:

\begin{align}\label{qstate decomp}
\begin{split}
| \psi \rangle \, = | \psi_{00} \rangle  + \, | \psi_{01} \rangle  + \, | \psi_{10} \rangle  + \, | \psi_{11} \rangle,
\end{split}
\end{align}
\\
where the components are defined by applying the respective projection operators:

\begin{align}\label{dec def}
\begin{split}
\begin{array} {l}
 | \psi_{00} \rangle \, =  (\boldsymbol{I} - \boldsymbol{A}) \,  (\boldsymbol{I} - \boldsymbol{B}) \,  | \psi \rangle, 
\\[3pt]
 | \psi_{01} \rangle \, =  (\boldsymbol{I} - \boldsymbol{A}) \, \boldsymbol{B} \, | \psi \rangle,
\\[3pt]
  | \psi_{10} \rangle \, = \boldsymbol{A} \, (\boldsymbol{I} - \boldsymbol{B}) \, | \psi \rangle,
\\[3pt]
 | \psi_{11} \rangle \, = \boldsymbol{A} \, \boldsymbol{B} \, | \psi \rangle .
 \end{array}
\end{split}
\end{align}
\\

We want to emphasize that the decomposition (\ref{qstate decomp}) is not in general a basis decomposition because vectors can live in a vector space of arbitrary dimension. 

We have :

\begin{align}\label{dec transf}
\begin{split}
 \boldsymbol{A} \, | \psi_{00} \rangle \, = 0   ~~,\, &~~  \boldsymbol{B} \, | \psi_{00} \rangle \, =  0 .
  \\
   \boldsymbol{A} \, | \psi_{01} \rangle \, = 0   ~~,\, &~~  \boldsymbol{B} \, | \psi_{01} \rangle \, =  | \psi_{01} \rangle.
   \\
  \boldsymbol{A} \, | \psi_{10} \rangle \, = | \psi_{10} \rangle  ~~,\, &~~  \boldsymbol{B} \, | \psi_{10} \rangle \, = 0.
  \\
 \boldsymbol{A} \, | \psi_{11} \rangle \, = | \psi_{11} \rangle  ~~,\, &~~  \boldsymbol{B} \, | \psi_{11} \rangle \, =  | \psi_{11} \rangle   .
\end{split}
\end{align}
\\

We then calculate the relationship between the probabilities and the square moduli of the various vectors   $| \psi_{00} \rangle$,  $| \psi_{01} \rangle$,  $| \psi_{10} \rangle$ and  $| \psi_{11} \rangle$  of the quantum state decomposition (\ref{qstate decomp}).  For an arbitrary state vector $| \chi \rangle$, we denote the square modulus $|\chi|^2\equiv\langle \chi | \chi\rangle $. Then we have the following relations:

\begin{align}\label{prob q logic}
\begin{split}
 \begin{array} {l}
 P_\psi(A)=|\psi_{10}|^2 + |\psi_{11}|^2,
 \\[3pt]
P_\psi(A\cap B)=|\psi_{11} |^2,
\\[3pt]
P_\psi(A\rightarrow B) =  |\psi_{01}|^2 + |\psi_{00}|^2 +  |\psi_{11}|^2 = 1 - |\psi_{10}|^2.
\\[3pt]
 \end{array}
\end{split}
\end{align}
\\

\section{Quantum-like Bayes' Conditions}\label{Quantum-like Bayes' conditions}

In order to verify whether equation (\ref{Quantum Bayes'}) is satisfied, we express the difference of the following probabilities:

\begin{align}\label{prob q logic}
\begin{split}
P_\psi(A) \, P_\psi(A\rightarrow B) - P_\psi(A\cap B) =
(| \psi_{10} |^2 + | \psi_{11} |^2) \, (1 - | \psi_{10} |^2) - | \psi_{11} |^2  \\ 
= | \psi_{10} |^2 - (| \psi_{10} |^2 + | \psi_{11} |^2) \,  | \psi_{10} |^2 \\
=  | \psi_{10} |^2 \, \big( 1 -  (| \psi_{10} |^2 + | \psi_{11} |^2) \big) .
\end{split}
\end{align}
\\

Equation (\ref{Quantum Bayes'}) is satisfied when the difference in equation (\ref{prob q logic}) is zero, two cases can occur.

\begin{itemize}
\item The first case is obtained when $ \,  | \psi_{10} |^2 = 0 \, $,
then we have $ \, P_\psi(A\rightarrow B) = 1 \, $ and the implication $ \, A\rightarrow B \, $
is certain. 

\item In the second case, the condition is realized if we have
$ \, | \psi_{10} |^2 + | \psi_{11} |^2 = 1 $ meaning that $ \, P_\psi(A) = 1 $. 
\end{itemize}

In other terms, 
the state $ \,   | \psi \rangle \, $ lies in the range of the projector $\boldsymbol{A}$.

In conclusion, the equality (\ref{Quantum Bayes'}) is satisfied if $ \,| \psi_{10} |^2 = 0 \, $ or $ \,|\psi_{01} |^2 = |\psi_{00} |^2=0\, $ . In the first case $ \, P_\psi(A\rightarrow B) = P_\psi(B | A)=1 \, $.

Using the previous notation, the classical Bayes relation takes the form:

\begin{align}\label{quantum-bayes}
\begin{split}
P_\psi(A) \,   P_\psi(B|A) = P_\psi(B) \,   P_\psi(A|B) .
\end{split}
\end{align}
\\

A natural choice is to establish the general conditions on the quantum state $|\psi \rangle$ in order to verify the preceding condition (\ref{quantum-bayes}), as previously we suppose that : 

\begin{align}\label{P(A)-not-zero}
\begin{split}
P_\psi(A) \equiv  | \psi_{10} |^2 + | \psi_{11} |^2  \not= 0
\end{split}
\end{align}
and

\begin{align}\label{P(B)-not-zero}
\begin{split}
P_\psi(B) \equiv  | \psi_{01} |^2 + | \psi_{11} |^2  \not= 0 .
\end{split}
\end{align}
\\

Moreover, the state $ | \psi  \rangle = | \psi_{00} \rangle  +  | \psi_{10} \rangle  +  | \psi_{01} \rangle  +  | \psi_{11} \rangle$
has a norm equal to 1 so we have:

\begin{align}\label{psi-norm-1}
\begin{split}
| \psi_{00} |^2 + | \psi_{10} |^2 + | \psi_{01} |^2 + | \psi_{11} |^2 = 1 .
\end{split}
\end{align}
\\
It satisfies the following two conditions:

\begin{align}\label{Bayes=Prob-Implication}
\begin{split}
P_\psi(B | A) \equiv   {{ P_\psi(A\cap B)}\over{P_\psi(A)}}   =  P_\psi(A\rightarrow B)
\end{split}
\end{align}
and

\begin{align}\label{Bayes=Prob-RevImplication}
\begin{split}
P_\psi(A | B) \equiv   {{ P_\psi(B\cap A)}\over{P_\psi(B)}}   =  P_\psi(B\rightarrow A) ,
\end{split}
\end{align}

the last two expressions constitute the quantum-like Bayes rule.
These two conditions take the algebraic form:

\begin{align}\label{Bayes-A}
\begin{split}
| \psi_{10} |^2 \, \big( 1 -  (| \psi_{10} |^2 + | \psi_{11} |^2) \big) = 0
\end{split}
\end{align}
and

\begin{align}\label{Bayes-B}
\begin{split}
| \psi_{01} |^2 \, \big( 1 -  (| \psi_{01} |^2 + | \psi_{11} |^2) \big) = 0 . 
\end{split}
\end{align}

If conditions (\ref{Bayes-A}) and (\ref{Bayes-B}) are satisfied, it is clear that relation (\ref{quantum-bayes}) is verified because we have on one hand $P_\psi(A)P_\psi(B|A)=P_\psi(A\cap B)$ and on the other $P_\psi(B)P_\psi(A|B)=P_\psi(A\cap B)$. In the following of this section, we explicit carefully the hypotheses
(\ref{P(A)-not-zero}), (\ref{P(B)-not-zero}), (\ref{Bayes-A}) and (\ref{Bayes-B}) that are necessary to establish this quantum-like Bayes relation. 

Four cases occur according to the product terms in the equations (\ref{Bayes-A}) and (\ref{Bayes-B}):
\begin{enumerate}
\item 
$ | \psi_{10} | = | \psi_{01} | = 0 $;

\qquad then $ P_\psi(A) = | \psi_{11} |^2 $ must be non-zero and the same for $ P_\psi(B) = | \psi_{11} |^2 $. In this case, the hypothesis take the form:

\begin{align}\label{hyp-1}
\begin{split}
| \psi_{10} | = | \psi_{01} | = 0 ,\, \,  | \psi_{11} |  \not = 0 .
\end{split}
\end{align}

\item 
 $ | \psi_{10} | = 0 \, $ and $ \, | \psi_{01} |^2 + | \psi_{11} |^2 = 1 $; 

\qquad then $ P_\psi(B) = 1 $  due to equation (\ref{P(B)-not-zero}). In consequence of 
(\ref{psi-norm-1}), we also have $ | \psi_{00} | = 0 $ and $ P_\psi(A) \not = 0 $
implies $ | \psi_{11} |   \not = 0 $. The hypothesis for this second case takes the complete form:

\begin{align}\label{hyp-2}
\begin{split}
| \psi_{10} | = | \psi_{00} | = 0 ,\,\,  | \psi_{01} |^2 + | \psi_{11} |^2 = 1 ,\,\,  | \psi_{11} |  \not = 0 .
\end{split}
\end{align}

\item
 $ | \psi_{01} | = 0 \, $ and $| \psi_{10} |^2 + | \psi_{11} |^2 = 1 $; 

\qquad this case is analogous to the previous one. We just have to exchange the roles of the two
operators $\boldsymbol{A}$ and $\boldsymbol{B}$ . We have $ P_\psi(A) = 1$ and $| \psi_{00} | =0$ as a consequence of the
normalization in~(\ref{psi-norm-1}). We must also have  $P_\psi(B) \not = 0 $ and in consequence we have 
$| \psi_{11} |   \not = 0 $. 
The hypothesis to get the relation (\ref{quantum-bayes}) takes the complete form in this third case:

\begin{align}\label{hyp-3}
\begin{split}
| \psi_{01} | = | \psi_{00} | = 0 ,\,\,  | \psi_{10} |^2 + | \psi_{11} |^2 = 1 ,\,\,  | \psi_{11} |  \not = 0 .
\end{split}
\end{align}

\item
 $ \, | \psi_{01} |^2 + | \psi_{11} |^2 = 1 \,$ and 
$ \,  | \psi_{10} |^2 + | \psi_{11} |^2 = 1 $; 

\qquad then due to  (\ref{psi-norm-1}), we have
$ | \psi_{00} | = | \psi_{01} | = | \psi_{10} | = 0 $ and we must have $| \psi_{11} | = 1 $.
In this fourth case, the hypotheses of the relation   (\ref{quantum-bayes})  can be written:

\begin{align}\label{hyp-4}
\begin{split}
| \psi_{11} | = 1 ,\,\,  | \psi_{00} | = | \psi_{01} | = | \psi_{10} | = 0 .
\end{split}
\end{align}
\end{enumerate}

\section{Application Examples on 2-Qubit States }
We will now discuss the application of the preceding conditions on some examples of 2-qubit quantum states. We write the general 2-qubit quantum state as a decomposition on an orthonormal basis $|ij \rangle$ ($i,j \in {0,1}$):

\begin{align}\label{qstate decomp 2-qubit}
\begin{split}
| \psi \rangle \, =c_{00} | 00 \rangle  + c_{01}| 01 \rangle  + c_{10}  |10 \rangle  + c_{11}| 11 \rangle .
\end{split}
\end{align}

In Eiegnlogic we choose that the conventional logical basis is the "z" basis; this is also the habit in the field of quantum information with the "computational basis" represented by qubits $| 0 \rangle$ and $| 1 \rangle$ and their compositions by the Kronecker product.

When the quantum state $|\psi \rangle$ is one of the basis states $| 00 \rangle$, $| 01 \rangle$, $| 10 \rangle$ or $| 11 \rangle$ we are in the classical logic case where probabilities transform into certainties or impossibilities and their values are given by the truth values given in table \ref{Truth table}. The resulting probabilities are given in table \ref{Eigensystem}.  The quantum-like Bayes rules (\ref{Bayes=Prob-Implication}) and (\ref{Bayes=Prob-RevImplication}) are here always satisfied.

\begin{table}[!h]
\caption{Probabilities for the quantum states in the eigensystem  $| 00 \rangle$, $| 01 \rangle$, $| 10 \rangle$  $| 11 \rangle$.}
\label{Eigensystem}
\begin{tabular}{|l|l|l|l|l|l|l|l|}
 \hline  
\small state &\tiny \( P(A) \)&\tiny \( P(B) \) &\tiny \( P(A \rightarrow B) \)&\tiny \( P(B \rightarrow A) \)&\tiny \( P(A)P(A \rightarrow B) \)&\tiny \( P(B)P(B \rightarrow A) \)&\tiny \( P(A \cap B) \)\\ \hline  

$| 00 \rangle$ & 0 & 0 & 1 & 1 & 0 & 0 & 0  \\ \hline  
$| 01 \rangle$ & 0 & 1 & 1 & 0 & 0 & 0 & 0  \\ \hline  
$| 10 \rangle$ & 1 & 0 & 0 & 1 & 0 & 0 & 0  \\ \hline  
$| 11 \rangle$ & 1 & 1 & 1& 1 & 1 & 1 & 1  \\ \hline 
\end{tabular}
\vspace*{-4pt}
\end{table}

Let's consider now examples where we have a uniform distribution of the coefficients over the basis states: $| c_{01} |=| c_{01} |=| c_{01} |=|c_{01} |=1/2$.
The uniform quantum state $|++\rangle$ is the tensor product of the two one-qubit states  $|+\rangle=\frac{1}{\sqrt{2}} (| 0 \rangle  + | 1 \rangle )$ and is a basis state of the Eigenlogic "x" system \cite{EL-Entropy 2020} and has the property of being a uniform superposition and a mutually unbiased basis (MUB) state:

\begin{align}\label{MUB state}
\begin{split}
|++\rangle = \frac{1}{\sqrt{2}} (| 0 \rangle  + | 1 \rangle ) \otimes \frac{1}{\sqrt{2}} (| 0 \rangle  + | 1 \rangle=\frac{1}{2} ( | 00 \rangle  +| 01 \rangle  +  |10 \rangle  + | 11 \rangle ),
\end{split}
\end{align}

for this state we have: $\, P_\psi(A) =P_\psi(B)= \frac{1}{4} + \frac{1}{4} = \frac{1}{2} \, $  , $P_\psi(A \rightarrow B)=P_\psi(B \rightarrow A)=\frac{1}{4} + \frac{1}{4} + \frac{1}{4}= \frac{3}{4}$ and  $P_\psi(A \land B)=\frac{1}{4} $.
\begin{table}[!h]
\caption{Probabilities for the uniform quantum states $|++\rangle$, $|+-\rangle$, $|-+\rangle$, $|--\rangle$. and the cluster state $|\psi_cl \rangle$ .} 
\label{MUB}
\begin{tabular}{|l|l|l|l|l|l|l|l|}
 \hline  
\small state &\tiny \( P(A) \)&\tiny \( P(B) \) &\tiny \( P(A \rightarrow B) \)&\tiny \( P(B \rightarrow A) \)&\tiny \( P(A)P(A \rightarrow B) \)&\tiny \( P(B)P(B \rightarrow A) \)&\tiny \( P(A \cap B) \)\\ \hline  
$|uniform\rangle$& $1/2$ & $1/2$ & $3/4$ & $3/4$ & $3/8$ & $3/8$ & $1/4$  \\ \hline  
\end{tabular}
\vspace*{-4pt}
\end{table}
In this case, we do not verify the Bayes' rule.

The same results are obtained for different phases in the coefficients. In particular for the other basis states of the "x" system $|+-\rangle$, $|-+\rangle$ and $|--\rangle$ where $|-\rangle=\frac{1}{\sqrt{2}} (| 0 \rangle  - | 1 \rangle )$ . This is also the case for the two-qubit \textit{cluster state} which is a maximally entangled quantum state:

\begin{align}\label{MUB state}
\begin{split}
| \psi_cl \rangle =\frac{1}{2}( | 00 \rangle  +| 01 \rangle  +  |10 \rangle  -| 11 \rangle ). 
\end{split}
\end{align}
\\

We see that the uniform states never satisfy the Bayes' quantum-like rule, the difference is $1/8$ for the probabilities of implication $A \rightarrow B$ and its reverse $B \rightarrow A$ .

Now let us combine states from the non-commuting "z" basis and "x" basis, such as for the states  $|0+\rangle$ and $|1+\rangle$:

\begin{align}\label{MUB state}
\begin{split}
|0+\rangle = | 0 \rangle  \otimes \frac{1}{\sqrt{2}} (| 0 \rangle  + | 1 \rangle)=\frac{1}{\sqrt{2}} (| 00 \rangle  + | 01 \rangle ),
\end{split}
\end{align}
\\
\begin{align}\label{MUB state}
\begin{split}
|1+\rangle =| 1 \rangle  \otimes \frac{1}{\sqrt{2}} (| 0 \rangle  + | 1 \rangle)=\frac{1}{\sqrt{2}} (| 10 \rangle  + | 11 \rangle ),
\end{split}
\end{align}
\\
or the states $|+0\rangle$ and $|+1\rangle$:

\begin{align}\label{MUB state}
\begin{split}
|+0\rangle = \frac{1}{\sqrt{2}} (| 0 \rangle  + | 1 \rangle) \otimes| 0 \rangle=\frac{1}{\sqrt{2}} (| 00 \rangle  + | 10 \rangle ),
\end{split}
\end{align}
\\
\begin{align}\label{MUB state}
\begin{split}
|+1\rangle =\frac{1}{\sqrt{2}} (| 0 \rangle  + | 1 \rangle)  \otimes  |1 \rangle =\frac{1}{\sqrt{2}} (| 01 \rangle  + | 11 \rangle ).
\end{split}
\end{align}
\\
\begin{table}[!h]
\caption{Probabilities for the mixed basis separable quantum states $|0+\rangle$, $|1+\rangle$,$|+0\rangle$, $|+1\rangle$.} 
\label{mixed basis}
\begin{tabular}{|l|l|l|l|l|l|l|l|}
 \hline  
\small state &\tiny \( P(A) \)&\tiny \( P(B) \) &\tiny \( P(A \rightarrow B) \)&\tiny \( P(B \rightarrow A) \)&\tiny \( P(A)P(A \rightarrow B) \)&\tiny \( P(B)P(B \rightarrow A) \)&\tiny \( P(A \cap B) \)\\ \hline  
$|0+\rangle $& $0$ & $1/2$ & $1$ & $1/2$ & $0$ & $1/4$ & $0$  \\ \hline
$|+0\rangle $& $1/2$ & $0$ & $1/2$ & $1$ & $1/4$ & $0$ & $0$  \\ \hline
$|1+\rangle $& $1$ & $1/2$ & $1/2$ & $1$ & $1/2$ & $1/2$ & $1/2$ \\ \hline
$|+1\rangle $& $1/2$ & $1$ & $1$ & $1/2$ & $1/2$ & $1/2$ & $1/2$  \\ \hline
\end{tabular}
\vspace*{-4pt}
\end{table}

In table \ref{mixed basis} we see that the states $|1+\rangle$ and $|+1\rangle$ satisfy the quantum-like  Bayes' rule.  For $|0+\rangle$ and $|+0\rangle$  the rule is partially satisfied for only one implication case, either $A \rightarrow B$ or $B \rightarrow A$,  for the other cases we have a difference of $1/4$ in the probabilities .

Let's now consider the entangled states of the \textit{Bell basis}:

\begin{align}\label{Bell states}
\begin{split}
\begin{array} {l}
| \Phi^+ \rangle \, =\frac{1}{\sqrt{2}} (| 00 \rangle  + | 11 \rangle ),
\\[9pt]
|\Phi^- \rangle \, =\frac{1}{\sqrt{2}} (| 00 \rangle  - | 11 \rangle ),
\\[9pt]
| \Psi^+ \rangle \, =\frac{1}{\sqrt{2}} (| 01 \rangle  + | 10 \rangle ),
\\[9pt]
| \Psi^- \rangle \, =\frac{1}{\sqrt{2}} (| 01 \rangle  - | 10 \rangle ).
\end{array}
\end{split}
\end{align}
\begin{table}[!h]
\caption{Probabilities for the Bell entangled quantum states.}
\label{Bell states}
\begin{tabular}{|l|l|l|l|l|l|l|l|}
 \hline  
\small state &\tiny \( P(A) \)&\tiny \( P(B) \) &\tiny \( P(A \rightarrow B) \)&\tiny \( P(B \rightarrow A) \)&\tiny \( P(A)P(A \rightarrow B) \)&\tiny \( P(B)P(B \rightarrow A) \)&\tiny \( P(A \cap B) \)\\ \hline  

$| \Phi^+ \rangle$ & $1/2$ & $1/2$ & $1$ & $1$ & $1/2$ & $1/2$ & $1/2$  \\ \hline  
$|  \Phi^- \rangle$ & $1/2$ & $1/2$ & $1$ & $1$ & $1/2$ & $1/2$ & $1/2$  \\ \hline  
$| \Psi^+ \rangle$ & $1/2$ & $1/2$ & $1/2$ & $1/2$ & $1/4$ & $1/4$ & $0$  \\ \hline  
$| \Psi^- \rangle$ & $1/2$ & $1/2$ & $1/2$& $1/2$ & $1/4$ & $1/4$ & $0$  \\ \hline 
\end{tabular}
\vspace*{-4pt}
\end{table}

In table \ref{Bell states} we see that there is agreement with the Bayes' rule for the states $| \Phi^+ \rangle$ and $| \Phi^- \rangle$ but not for $| \Psi^+ \rangle$ and $| \Psi^- \rangle$.  So we see that entanglement is not a discriminating condition.

A last interesting fact is that the linear form involving logical implication probabilities, given in equation (\ref{proba mat implic 4}), is always verified for all the quantum situations discussed above. 

\section{Discussion}

This paper presents Eigenlogic as a framework to unify Bayesian and Born probabilities within a common formalism. By representing logical propositions as projection operators, we construct a probabilistic logic that is consistent with quantum measurement principles, allowing for inference in both classical and quantum domains.

In essence, while Bayes theorem operates on classical probabilities, the Born rule governs probabilities in quantum systems. The Bayes theorem updates the probabilities based on new information, and the Born rule updates the probabilities based on measurement, with both sharing a similar updating role in their respective frameworks. Eigenlogic offers a conceptual bridge between these two by viewing projection operators in quantum mechanics as analogs of conditional probabilities.

In the context of this work we want to quote several recent researches regarding probabilistic and quantum inference methods.

Andrei Khrennikov initiated the so-called \textit{quantum-like} approach \cite{Khrennikov 2010, Khrennikov 2018} using the mathematical framework of quantum mechanics to tackle problems outside physics, principally in the social sciences \cite{Haven Khrennikov 2013} (see also Michel Bitbol \textit{et al.} in \cite{Bitbol 2016}). 
 The method presented in this paper based on Eigenlogic can be considered a quantum-like approach \cite{EL-Entropy 2020}.
 
Quantum-like modeling has been used to explore \textit{quantum cognition} problems \cite{Busemeyer Bruza 2012}, where quantum probability theory can better explain cognitive processes, particularly in situations where classical Bayesian probability theory fails, as for the following non-classical effects: \textit{order effect}, \textit{cognitive dissonance} and \textit{zero prior trap}.

In Bayesian updating the problem of \textit{zero prior trap}  is encountered when an event has a zero prior probability. In this case no amount of new evidence can update this probability because Bayes rule (\ref{Bayes rule}) involves division by the prior probability. This is problematic in real-world cognitive processes where individuals might initially ignore or be unaware of an event but later recognize its relevance. According to the results presented in this paper, one could use the alternative equation (\ref{proba mat implic 4}), given in section \ref{Probabilistic Inference}, instead of the Bayes rule (\ref{Bayes rule}) in order to avoid \textit{zero prior trap} because the equation is now linear with no division operation.

Among other methods inspired by quantum mechanics, we have \textit{Positive-Operator Valued Measures} (POVMs) and \textit{Projection-Valued Measure}  (PVM) in cognitive modeling \cite{Ozawa Khrennikov 2020}, which is a powerful approach for explaining the effects of contextuality, uncertainty, and non-classical probability in decision making, psychology, and artificial intelligence. Our method based on Eigenlogic projection operators could also benefit this research.

\textit{Quantum Bayesianism} (QBism) \cite{Fuchs 2010} represents a departure from traditional interpretations of quantum mechanics, emphasizing a personalist Bayesian perspective where quantum probabilities reflect the subjective degrees of belief of an agent. The concept of \textit{intersubjectivity} in quantum mechanics has been introduced \cite{Ozawa 2019,Khrennikov 2024} in order to explore whether the assignments of quantum states of different agents must eventually converge; this challenges QBism's claim that quantum states are purely subjective beliefs.

We can also cite the \textit{Paraconsistent Logic} framework \cite{da Costa 2007} which offers significant advantages in addressing the logical inconsistencies of quantum mechanics. For example, the excluded middle principle (which states that every proposition is either true or false) is often problematic in quantum systems due to phenomena like superposition and entanglement.

Also recent classical approaches for logical inference have been developed to unify logic-based inference with probabilistic updating. These models include: \textit{Probabilistic Conditional Logic} \cite{Goertzel 2009} that extends classical logic by assigning probabilities to conditionals; \textit{Bayesian Networks with Logical Constraints} \cite{Han Pereira 2013} which use directed acyclic graphs and \textit{Non-Monotonic Logics and Probability} \cite{Fields Glazebrook 2020} where non-monotonic reasoning adjust beliefs dynamically via Bayesian updates.

All these models enhance traditional Bayesian inference by incorporating context awareness, hierarchical structuring, stochastic logic, and dynamic updates. They have broad applications in AI, robotics, cognitive science, decision theory, and human-computer interaction.

Our work demonstrates that classical inference rules such as \textit{modus ponens} can be extended to quantum contexts.
As quantum computing and machine learning continue to develop, frameworks such as Eigenlogic that harmonize probabilistic reasoning in quantum systems will become increasingly relevant.

This paper could motivate further exploration of logical inference mechanisms within quantum information processing and points to potential innovations in quantum learning systems that operate under conditions of uncertainty.

\smallskip \smallskip \smallskip \smallskip \smallskip \noindent 
{\bf \Large{Acknowledgments}}

\noindent The authors would like to thank the referees for their detailed reading and interesting remarks.

\end{document}